\documentclass{article}
\usepackage{amsmath,amsthm}
\usepackage{amssymb,latexsym}
\usepackage[mathscr]{eucal}
\usepackage{setspace}
\usepackage{graphics}
\usepackage{array}

\newcommand{\pa}{\partial}

\newcommand{\al}{\alpha}

\newcommand{\rp}{\right)}
\newcommand{\lp}{\left(}
\newcommand{\rb}{\right]}
\newcommand{\lb}{\left[}
\newcommand{\rel}{\right\}}
\newcommand{\lel}{\left\{}

\newcommand{\beq}{\begin{equation}}
\newcommand{\eq}{\end{equation}}

\newcommand{\bfv}{{\bf v}}

\newcommand{\na}{\nabla}

\newcommand{\barray}{\begin{array}}
\newcommand{\earray}{\end{array}}

\usepackage{graphicx}
\graphicspath{{converted_graphics/}}
\begin{document}

\title{\textbf{Rayleigh-Taylor Instability in a Compressible Fluid}}         
\author{B.K. Shivamoggi\\
University of Central Florida\\
Orlando, FL 32816-1364\\
}        
\date{}          
\maketitle

\large{\bf Abstract}

Rayleigh-Taylor instability in a compressible fluid is reconsidered. The density is allowed to vary with pressure under the barotropy assumption. For the case with equal speeds of sound in the two superposed fluids, in order to give a non-trivial compressibility correction to the Rayleigh-Taylor growth rate, the compressibility correction is calculated to $O(g^2/k^2a^4)$. To this order, compressibility effects are found to reduce the growth rate. 

\pagebreak

\section{Introduction}      

The instability of the interface between two fluids having different densities and accelerated towards each other is called the Rayleigh-Taylor instability (Rayleigh \cite{Ray}, Taylor \cite{Tay}). For the case of two superposed uniform fluids separated by a plane interface, the amplitude of a small disturbance which is periodic in the horizontal interfacial plane will vary in time like $e^{nt}$, where

\beq
n^2=\frac{\rho_2 -\rho_1}{\rho_2+\rho_1} gk
\eq
\vspace{0.15in}

\noindent
while $\rho_1$ and $\rho_2$ are, respectively, the densities of the lower and the upper fluids, $k$ the horizontal wavenumber of the disturbance, and $g$ the gravitational acceleration. (1) shows that if $\rho_2>\rho_1$, the interface is unstable. Chandrasekhar \cite{Cha} and Hide \cite{Hid} included the effect of viscosity while Bellman and Pennington \cite{Bel} showed that the effect of surface tension $T$ at the interface in the linear problem is to produce a critical wavenumber $k_c$, given by 

\beq
k_c=\lb \frac{g(\rho_2-\rho_1)}{T}\rb^{\frac{1}{2}}
\eq 
\vspace{0.15in}

\noindent so the interface is unstable or stable according to whether the wavenumber $k$ is less than or greater than $k_c$. Rayleigh-Taylor instability plays a crucial role in inertial confinement fusion (Petrasso \cite{Pet}) and gravitational fusion in stars (Burrows \cite{Bur}). 

The effect of compressibility on Rayleigh-Taylor instability was considered by Vanderwoort \cite{Van} and Plesset and Hsieh \cite{PH} but the results were in disagreement with each other (Shivamoggi \cite{Shi}). Several attempts have since been made to clarify the role of compressibility (Plesset and Prosperetti \cite{Ple}, Bernstein and Book \cite{Ber}), but the picture is not completely clear. In view of the inconsistency, as noted by Plesset and Prosperetti \cite{Ple}, in assuming the constancy of both density and speed of sound $a$ by Vanderwoort \cite{Van}, in this paper, we reformulate Vanderwoort's \cite{Van} development by relaxing the constant density assumption and allow density to vary with pressure under the barotropy assumption. The resulting dispersion relation turns out to be different from the one given by Plesset and Prosperetti \cite{Ple} in general, but agrees with the latter to $O(g/ka^2).$ However, the latter compressibility correction disappears when the speeds of sound in the two fluids become equal. In order to get a non-trivial result for the latter case, in this paper we proceed further and give the compressibility correction to next order, namely $O(g^2/k^2a^4)$.

\section{Governing Equations}

Consider the static state of infinite fluid arranged in horizontal layers in a uniform gravitational field $y$, the density $\rho$ and the pressure $p$ are functions of the vertical coordinate $z$ satisfying the equation of hydrostatic equilibrium -

\beq
\frac{dp}{dz}=-\rho g.
\eq

\noindent To determine the character of this equilibrium, consider the evolution of small disturbances $\delta \rho$ in $\rho$, $\delta p$ in $p$ and $\bfv$ in velocity on this static equilibrium state. The linearized equations governing these disturbances are - 

\vspace{0.15in}

continuity:\begin{equation}\frac{\pa\delta\rho}{\pa t} + \bfv\cdot\nabla\rho +\rho\nabla\cdot\bfv=o\\
\end{equation}

motion:\begin{equation} \rho \frac{\pa \bfv}{\pa t}=-\na\delta p-g\delta \rho~ {\hat{\bf i}_z}
\end{equation}

\noindent
and following Vanderwoort \cite{Van}, we use that barotropic condition (this is what indeed Vanderwoort \cite{Van} used without explicitly mentioning it) relating $p$ and $\rho$ - 

\beq
\frac{\pa\delta p}{\pa t} + \bfv\cdot\na p = a^2 \lp \frac{\pa \delta \rho}{\pa t} + \bfv\cdot\na\rho\rp
\eq

\noindent
$a$ being the speed of sound. 

Assuming the disturbances to vary with $x,y$ and $t$ according to $e^{i(k_1 x+k_2 y)+nt}$ we obtain (Vanderwoort \cite{Van}) - 

\beq
n^2 D\lb\frac{\rho}{(k^2+n^2/a^2)} Dw\rb - n^2 \rho w + gk^2 \lb\lp D+\frac{g}{a^2}\rp\lel\frac{\rho}{(k^2+ n^2/a^2)}\rel\rb w=o
\eq

\noindent
where,

\beq
\bfv\equiv<u,v,w>\ \ and\ \ D\equiv\frac{d}{dz}.\notag
\eq

The relevant boundary conditions at an interface supporting discontinuities in $\rho$ and $a$ are the kinematic jump condition -

\beq
\Delta w = o
\eq

\noindent and the dynamic jump condition that follows from equation (7) (Vanderwoort \cite{Van}) 

\beq
n^2 \Delta \lb\frac{\rho}{(k^2+ n^2/a^2)}Dw\rb + gk^2\Delta \lb\frac{\rho}{(k^2+ n^2/a^2)}\rb(w)_o=o.
\eq

\section{The Rayleigh-Taylor Configuration}

Consider now the case of two semi-infinite fluids separated by a horizontal plane interface at $z=o$ with the pressure gradient in each fluid given by equation (3). Following Vanderwoort \cite{Van}, we assume the speed of sound in the two fluids to be constant so that $p$ is a linear function of $\rho$. However, the density under the barotropy assumption coupled with equation (3) in each fluid is now given by (Mathews and Blumenthal \cite{Mat})

\beq
\frac{1}{\rho} D\rho = -\frac{g}{a^2}.\
\eq
\vspace{0.15in}
\noindent It may be noted that the constant density assumption coupled with the constant sound speed assumption used by Vanderwoort \cite{Van}, as noted by Plesset and Prosperetti \cite{Ple}, is inconsistent. 

Using equation (1o), equation (7) becomes

\beq
D^2 w- \lp\frac{g}{a^2}\rp Dw - \lp k^2+\frac{n^2}{a^2}\rp w = o
\eq

\noindent
which agrees with the one given by Plesset and Prosperetti \cite{Ple}. 

One may seek a solution to equation (11) of the form
\beq
w\sim e^{qz}
\eq

\noindent
which then leads to 

\beq
q^2-\lp\frac{g}{a^2}\rp q- \lp k^2+\frac{n^2}{a^2}\rp=o
\eq

\noindent
from which,

\beq
q_{1,2} = \frac{g}{2a^2} \pm \sqrt{\frac{g^2}{4a^4}+\lp k^2 + \frac{n^2}{a^2}\rp.}
\eq

Assuming the disturbances to be bounded at infinity and invoking the jump condition (8), we obtain

\beq
w=\lel\begin{array}{ll}
w_o e^{q_1 z} ,z<o\\
w_o e^{q_2 z} ,z>o.
\end{array}
\right.
\eq

Using (15), the jump condition (9), gives the dispersion relation -

\beq
n^2\lb \lp\frac{\rho_2}{k^2 + n^2/a_2^2}\rp q_2 - \lp\frac{\rho_1}{k^2 + n^2/a_1^2}\rp q_1 \rb + gk^2 \lb\frac{\rho_2}{k^2 + n^2/a_2^2}
 - \frac{\rho_1}{k^2 + n^2/a_1^2}\rb =o.
\eq

\noindent Here, subscripts 1 and 2 refer to the upper and the lower fluids, respectively. 

\bigskip
Introducing

\beq
\mu\equiv\frac{q}{k},\ \al\equiv\frac{\rho}{1+n^2/k^2a^2}
\eq

\noindent
(16) may be rewritten as 

\beq
n^2=\frac{gk(\al_2 - \al_1)}{\al_1\mu_1 - \al_2\mu_2}.
\eq

It may be noted that the dispersion relation given by Plesset and Prosperetti \cite{Ple} does not agree with (18) - in the present notation, the one given by \cite{Ple} is 

\beq
n^2 = \frac{gk (\rho_2 - \rho_1)}{\rho_1/\mu_1 - \rho_2/\mu_2}
\eq

\noindent
which is different from (18). However, (19) turns out to agree with (18) to $O\lp g/ka^2\rp.$

\section{Small Compressibility Limit}

The complexity of (16)\,(or (18)) makes physical interpretations harder without a numerical calculation. It is useful to consider (16) in the small compressibility limit - $\lp g/ka^2\rp\ll 1$.

In this limit, on noting from (17) that

\begin{align}
\mu_{1,2} & = \frac{g}{2ka_{1,2}^2}\pm \sqrt{\frac{g^2}{4k^2a_{1,2}^4} + \frac{n^2}{k^2a_{1,2}^2} + 1}\notag\\
\end{align}
\noindent or
\begin{align}
\mu_{1,2} & = \frac{g}{2ka_{1,2}^2}\pm\lp 1 + \frac{n_O^2}{2k^2a_{1,2}^2} \rp + O\lp \frac{g^2}{k^2 a_{1,2}^4}\rp
\end{align}

\noindent
where $n_0$ is the growth rate in the incompressible limit $\lp\lp g/ka^2\rp\rightarrow 0\rp,$

\beq
n_0 = \sqrt{gk\lp\frac{\rho_2 - \rho_1}{\rho_2 + \rho_1}\rp},
\eq

\noindent
we obtain from (18),

\beq
n^2 = n_0^2 \lb 1+ \frac{\rho_1\rho_2 g}{k(\rho_1+\rho_2)^2} \lp\frac{1}{a_1^2}- \frac{1}{a_2^2}\rp + O\lp\frac{g^2}{k^2a_{1,2}^4}\rp\rb
\eq
\noindent
which agrees with the one given by Plesset and Prosperetti \cite{Ple}.

In the special case $a_1 = a_2$, (23) implies that compressibility effects disappear from this problem to $O(g/ka^2).$ We now need to proceed further and include in (23) terms to $O(g^2/k^2 a^4)$.

First, on noting from (17) that 

\beq
\mu_{1,2} = \frac{g}{2ka_{1,2}^2} \pm \lb 1 + \frac{n_0^2}{2k^2 a_{1,2}^2} + \frac{1}{8k^4 a_{1,2}^4} (g^2 k^2 - n_0^4)\rb + O\lp\frac{g^3}{k^3 a_{1,2}^6}\rp
\eq

\noindent
we obtain from (18), with $a_1 = a_2 = a$,

\beq
n^2 = n_0^2 \lb 1 - \frac{g^2}{2k^2 a^4}\lp\frac{\rho_1\rho_2}{(\rho_1 + \rho_2)^2}\rp + O\lp\frac{g^2}{k^3 a^6}\rp\rb.
\eq

\noindent
(25) shows that, in the special case $a_1 = a_2$, compressibility effects have a stabilizing character on this problem - in this special case, the general belief that fluid compressibility provides a sink for the potential energy released at the interface appears to be valid.

\section{Discussion}
In this paper, we have reconsidered Rayleigh-Taylor instability in a compressible fluid. The density is allowed to vary with pressure under the barotropy assumption. For the case with equal speeds of sound in the two superposed fluids, in order to give a non-trivial compressibility correction to the Rayleigh-Taylor growth rate, we have calculated the compressibility correction to $O\lp g^2/k^2 a^4\rp.$ To this order, compressiblity effects are found to reduce the growth rate. In this special case of equal speeds of sound, the fluid compressibility appears to provide a sink for the potential energy released at the interface.

\section{Acknowledgements}

This work was carried out during the author's visit to the Los Alamos National Laboratory.

\end{document}